\documentclass[twoside]{book}
\usepackage{graphicx}

 \makeatletter
\ifx\UNDEF\mail\def\mail{ }\else\fi
\ifx\UNDEF\prange\def\prange{0 0}\else\fi

\gdef\@empty{}
\def\Mail#1 #2 {\gdef\thecontact{#1}\gdef\theaddr{#2}}
\def\Range#1 #2 {\gdef\thefirstpage{#1}\gdef\thelastpage{#2}}
{\let\'\mail \expandafter\Mail\' }  
{\let\'\prange \expandafter\Range\' }   
 \gdef\@shtitle{\relax}
 \long\def\shtitle#1{\gdef\@shtitle{#1}}
 \long\def\author#1{\gdef\@author{#1}}
 \def\affil#1{\par\noindent{\rm#1\par}}
 \gdef\@abstract{}
 \long\def\abstract#1{\gdef\@abstract{#1}}

 \def\maketitle{\thispagestyle{empty}\chapter{\@title}}
 \renewcommand\chapter{\if@openright\cleardoublepage\else\clearpage\fi
                    \thispagestyle{empty}%
                    \global\@topnum\z@
                    \@afterindentfalse
                    \secdef\@chapter\@schapter}
 \def\@makechapterhead#1{%
  \vspace*{50\p@}%
  {\parindent \z@ \raggedleft \normalfont
    \ifnum \c@secnumdepth >\m@ne
      \if@mainmatter
        \par\nobreak
        \vskip 20\p@
      \fi
    \fi
    \interlinepenalty\@M
    \Huge \bfseries #1\par\nobreak
    \vskip.25in
    \large\bfseries\@author\par\nobreak
    \vskip 40\p@}
    \ifx\@abstract\@empty\else{\small\@abstract\par\vskip20\p@}\fi
  }

\DeclareRobustCommand\em
        {\@nomath\em \ifdim \fontdimen\@ne\font >\z@
                       \upshape \else \slshape \fi}

\def\@begintheorem#1#2{\sl \trivlist \item[\hskip \labelsep{\bf #1\ #2}]}
\def\@opargbegintheorem#1#2#3{\sl \trivlist
     \item[\hskip \labelsep{\bf #1\ #2\ (#3)}]}

  \def\@arabic#1{\number #1} 

\long\def\@makecaption#1#2{
    \vskip\abovecaptionskip
    \sbox\@tempboxa{{\small {\bf #1}: #2}}%
    \ifdim\wd\@tempboxa>\hsize
        {\small {\bf #1}: #2\par}
    \else
       \global\@minipagefalse
       \hbox to\hsize{\hfil\box\@tempboxa\hfil}
    \fi
    \vskip \belowcaptionskip}

\def\figstrut#1{\hbox to\linewidth{\vrule height#1\hfill}}

\makeatother

\shtitle{Statistical properties of agent-based market area model}
\title{Statistical properties of agent-based market area model}
\author{Zolt\'an Kuscsik, Denis Horv\'ath\affil{ Department of Theoretical Physics and Astrophysics, 
              University of P.J. \v{S}af\'arik, Ko\v{s}ice, Slovak Republic   }}
\abstract{
One dimensional stylized model taking into account spatial activity of firms with uniformly
distributed customers is proposed.  The spatial selling 
area of each firm is defined by a short interval
cut out from selling space (large interval). In this representation,
the firm size is directly associated with the size of
its selling interval.

 The recursive synchronous dynamics of economic evolution
is discussed where the growth rate is proportional to the firm size incremented 
by the term including the overlap of the selling area with areas 
of competing firms. Other words, the overlap of selling areas inherently generate
a negative feedback originated from the pattern of demand.
Numerical simulations focused
on the obtaining of the firm size distributions uncovered that
the range of free parameters where the Pareto's law holds
corresponds to the range for which the pair correlation
between the nearest neighbor firms attains its minimum.
}
\begin{document}
\maketitle
\section{Introduction}\label{Int}

The study of elemental interactions in social and economical systems 
has a great importance to understand the large-scale system properties. 
One of the universal large-scale properties exhibited by social systems 
in a robust way is the {\it Pareto's law} of wealth distribution and firm 
size \cite{Clementi2006,Hegyi2007,Sornette2007}. Pareto's law 
is generally associated to the observation, that personal income of individuals, 
the size of companies are distributed by power-law. 

The formation of power-laws has generally complex origin. Among other approaches, 
highly sophisticated multi-agent
models have been developed~\cite{Bouchaud2000,Rawlings2004,samanidou-2007} 
to explain the power-laws observed in various social systems.

In this paper we propose agent-based model that emphasizes role 
of spatial location of firm within the limited market area. The model approximates 
the basic mechanisms of competency that simply follows from the spatial positions and selling
activities of firms.

An extensive economic literature exists that deals with the competitiveness 
as consequence of location. One dimensional model of spatial duopoly 
introduced by Hotelling~\cite{Hotelling1929} has assumed that consumers 
are continuously and uniformly distributed along a line segment.
The model of firm distribution in a non-uniform environment has 
been developed by Lawrence~\cite{Lawrence1975}. It predicts 
firm density in an urban setting in which the population density 
decreases exponentially with the distance from the center. 
Erlenkotter~\cite{Erlenkotter1989} has considered uniformly 
distributed 
demand over the infinite plane. He has discussed various regular 
two-dimensional market shapes. An elegant and advanced 
multistore competitive model of two firms in a finite business area 
has been introduced by Dasci and Laporte~\cite{dasci2005}. 
This model has been investigated for one and two dimensional 
geographical markets. It assumes the costumers are dispersed through 
space in only one direction along some coordinate $x\in(0,1)$. 
In this regard it is useful to mention the functional 
expression for total revenue per firm
\begin{equation}
\int_{0}^{1}\, Q(x) f(x) \rm{d}x\,,  
\end{equation}
where $f(x)$ is the probability density function of the customers 
multiplied by the probability $Q(x)$ that customer at $x$ 
patronizes product of given firm.

As we have mentioned before, our present approach also pays attention to spatial aspects. The approach comes 
from ecologic-economic feedback concept of regulated factory emissions introduced by us recently~\cite{Kuscsik2007}. 
The work points out an emergence of critical properties in a two dimensional system 
with spatially distributed agents balancing the conflicting objectives. The model assumes 
that sources of diffusive emissions compete with the distributed sensorial agents. 
The analysis of the complex numerical data yield us to reductionist and 
purely geometric formulation that
is related to coverage percolation problem. The geometric idea has 
been applied here to study the spatial distribution of the competitive firms 
reduced to basic geometric objects that cover market area. Our stylized spatial model 
is defined as it follows.

\section{Firm growth}

We assume that each firm acts as a seller agent 
of a product from the same sort of industry or it
behaves as a provider of some service business. 
The spatial economic activity of the $i$th agent is 
defined by its position $x_i^{(t)}\in(0,L)$ 
and by its selling area 
$(x_i^{(t)}-r_i^{(t)},x^{(t)}_i +r_i^{(t)})_{\mbox{mod} L}$, 
where $L$ is the constant size of one dimensional market space 
with periodic boundary conditions and 
$r_i^{(t)}$ is the selling radius of firm $i$. 
It should be noted that radius does 
not mean strictly the space of the physical activity 
of the seller but it can be understood as a radius up 
to which the customers are attracted. We have considered 
customers uniformly distributed along a straight line. 
Interestingly, such arrangement is typical 
for the restaurants distributed along a main road or highway~\cite{dasci2005}. 

We follow with definition of the measure 
of the spatial activity of the $i$th firm
\begin{equation}
s_i^{(t)}(x) = \left\{ 
\begin{array}{cl}
1 & \mbox{if $i$ claims to sell at the position $x$} \\
\\
0 & \mbox{if $x$ is not from the agent's selling area} 
\end{array}\right.
\end{equation}
 
Generally, large selling area means more potential consumers 
covered by delivery of given product which results  
higher interest connected to higher profit. Without negative economic feedback, 
the continuous investment of the constant fraction of income yields 
to the exponential growth of the firm size and its sale. 
Several facts that yield the negative feedback between 
firm and its environment should be mentioned:  
(i)  the transportation costs of products are convex functions
     of distance~\cite{PhilipMcCann06012005}; 
(ii) the complexity of firm management 
     grows with a firm size 
(iii) larger firms use 
      more sophisticated and thus more expensive 
      information technologies; 
(iv) the presence of two or more competitive products 
     in the same location affects the prices as well as the annual 
     sales. 

Here we assume only a negative feedback that originates 
from the spatial {\it overlap} of selling areas. 
The overlap of $i$th firm area with 
the areas of the remaining $(N-1)$ firms is 
defined by 
\begin{equation}
\Omega_i^{(t)} = 
\sum_{
\stackrel{j=0}{j\neq i}
}^{N} 
\int_0^{L} s_i^{(t)}(x) 
	s_j^{(t)}(x) 
{\rm d} x \,.
\end{equation}

With this firm-firm interaction picture in mind, 
we suggested the dynamical rule of the firm growth 
\begin{equation}
r_i^{(t+1)} = r_i^{(t)} + \alpha  r_i^{(t)} - 
\beta \Omega_i^{(t)}\,,
\end{equation}
where $\alpha>0$ and $\beta>0$ are constant parameters 
controlling the instantaneous growth. The term 
$-\beta \Omega_i^{(t)}$ can be interpreted 
as a negative feedback that reflects the competition. The {\it selling area} of firm $i$ is expressed by 
\begin{equation}
S_i^{(t)} = \int_{0}^{L}\, s_i^{(t)}(x)\, 
{\rm d} x = 2 r_i^{(t)}\,.
\end{equation}

\section{Firm establishment and bankruptcy}

In the stylized version of the model we study the 
firm is established at random position 
with a small random initial selling radius 
$r_i^{(t)} \in (r_{\rm a}, r_{\rm b})$, 
where $r_{\rm a}$ is assumed to be the lower bound 
of profitability (smallest firm). Therefore, the bankruptcy 
of a firm occurs when $r_i^{(t)}<r_{\rm a}$. At the same time new 
firm (with the same index) is established at a new random position with some initially 
random size. This death-birth process is analogous to the so called extremal 
dynamics principle~\cite{Boettcher2001} applied to e.g. models of the wealth 
distribution~\cite{Pianegonda2003} and 
stock markets~\cite{Horvath2007}.
 
\section{Simulation and Numerical results}

The choice of parameters is chronic problem specially in models of 
the social-economic systems. 
Although models of interacting agents give qualitative predictions that in
many aspects resemble behavior of real-world systems, in the most cases, 
the quantitative analysis needs laborious tuning of parameters until the range
is reached for which the phenomenon of interest takes place. 
Our simulations were performed with constant number of firms $N=500$ for 
predefined market area  $L=3 \times 10^5$.  The constant growth factor 
$\alpha=0.01$ and the initial range of the selling space constrained 
by radii $r_{\rm a} = 2.0$ and $r_{\rm b} = 5.0$ is chosen to invoke 
steps much smaller than $L$.

To reach the stationary regime $10^5$ the initial synchronous system 
updates were discarded. The information from subsequent $10^6$ updates 
spent in stationary regime has been recorded. Their analysis has 
uncovered that firm-firm interaction controlled by $\beta$ 
admits to establish market regimes that differ in size distributions. 
We observed that sufficiently large $\beta$ leads to the 
market with lowered overlaps. On other hand, sufficiently small 
$\beta$ supports formation of oligopoly that cover a dominant area 
of available market space. The important for us power-law distributions 
are observable only for exceptional $\beta$~(see fig~\ref{fig3}). 
This finding opens a question: what regulatory real-world economic principle 
controls the sustaining of the empirically relevant power-law regimes. 
The related question is the optimization of free parameters.  
For this purpose we examined several heuristic criteria. 
The most attractive seems to be an extremal entropy 
principle~\cite{Raine2006}, but in that case one faces to 
the usual problem of the proper entropy definition. 

More pragmatic, however, less fundamental attempt comes from our 
analysis of firm-firm correlations. For this purpose the pair 
correlation function 
$C$ of sizes of nearest neighboring firms ($k$) 
at positions $x_i^{(t)} \leq x_k^{(t)}$
can be defined as
\begin{equation}
\label{correlation}
C = 
\Bigg< 
\frac{
\frac{1}{N} \,  
\sum_{i=1}^N r_i^{(t)} r_k^{(t)}   -   
\left(   
\frac{1}{N} 
\sum_{i=1}^N  r_i^{(t)} \right)^2  
} { 
\frac{1}{N} \,   \sum_{i=1}^N ( r_i^{(t)} )^2 - 
\left(    \frac{1}{N}  \sum_{i=1}^N  r_i^{(t)} \right)^2   
}
\Bigg>_t  \,.
\end{equation}
Here $\langle\ldots \rangle_t$ denotes temporal stationary average.  
The calculations for different $\beta$ uncovered that minimum of 
correlation function corresponds to parameter (or narrow interval of parameters) 
for which nearly power-law distributions can be fitted quite well. 
This extremal principle reflects the possible importance of the measurements of the 
spatial correlations in social and economic systems.

\begin{figure}
\centerline{
\includegraphics[width=0.7\textwidth]{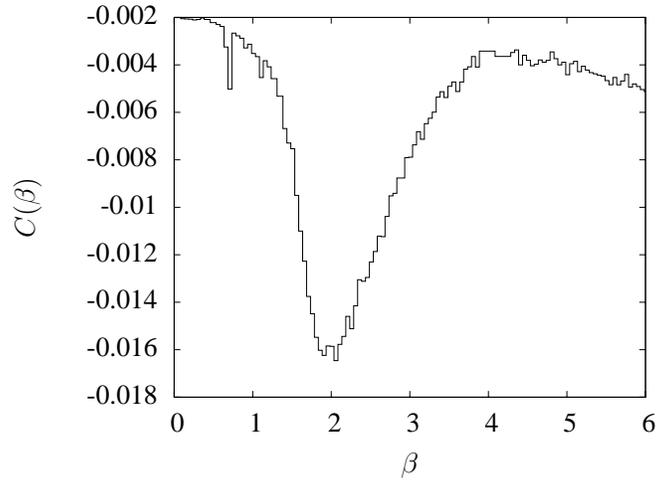}}
\caption{
Plot of the pair correlation function defined by Eq.~\ref{correlation} 
as a function of $\beta$.  The Pareto's law for  the 
firm size belongs to $\beta$ where $C(\beta)$ attains 
its minimum. } 
\label{fig1}
\end{figure}
	
\begin{figure}
\centerline{
a.)\includegraphics[width=0.5\textwidth]{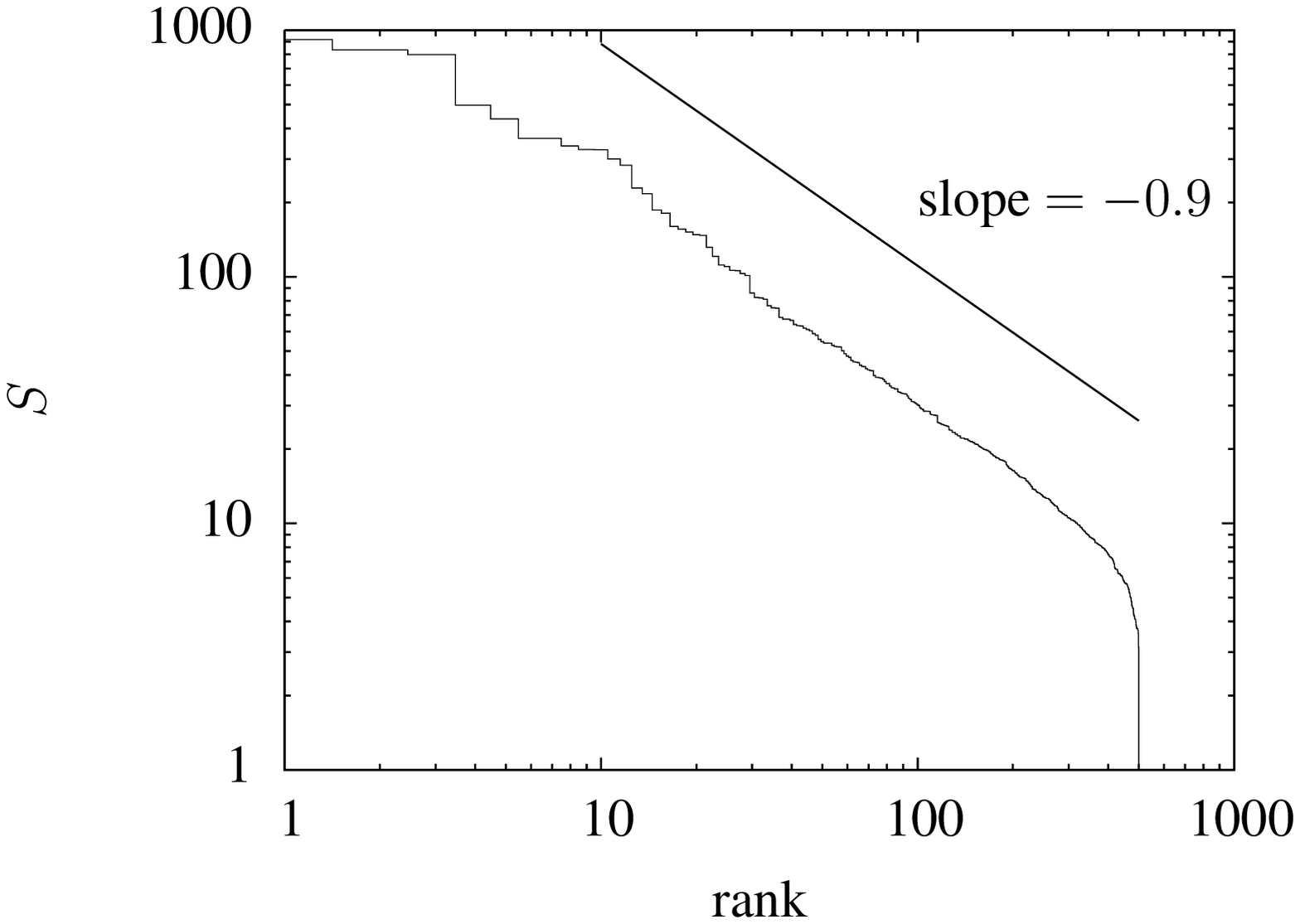}
b.)\includegraphics[width=0.5\textwidth]{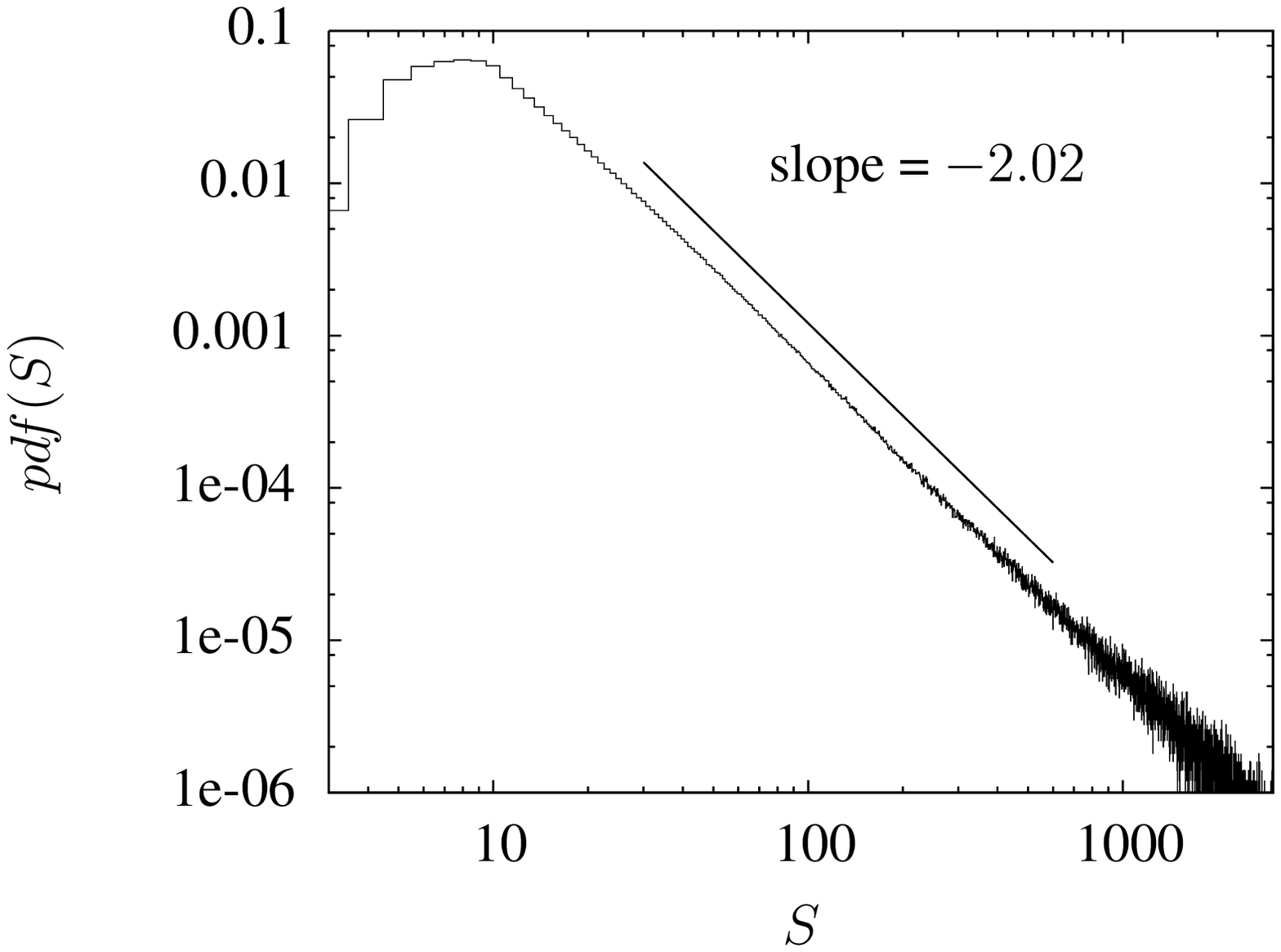}
}
\caption{
a.) Plot of sizes of firms as a 
function their rank for $\beta=2.0$.  
The fitted power-law index is close to unity. b.) 
The distribution of firm sizes. }
\label{fig2}
\end{figure}

\begin{figure}
\centerline{
\includegraphics[width=0.8\textwidth]{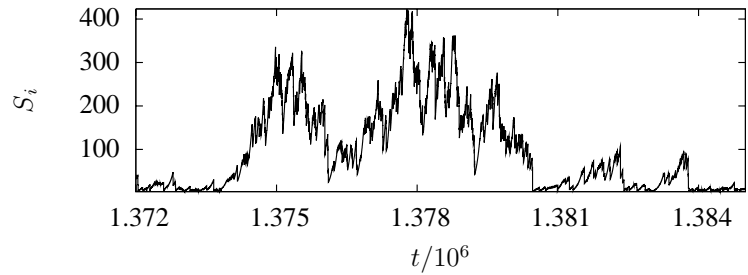}}
\caption{The time evolution of size of selected firm.}
\label{fig4}
\end{figure}

\begin{figure}
\centerline{
\includegraphics[width=0.8\textwidth]{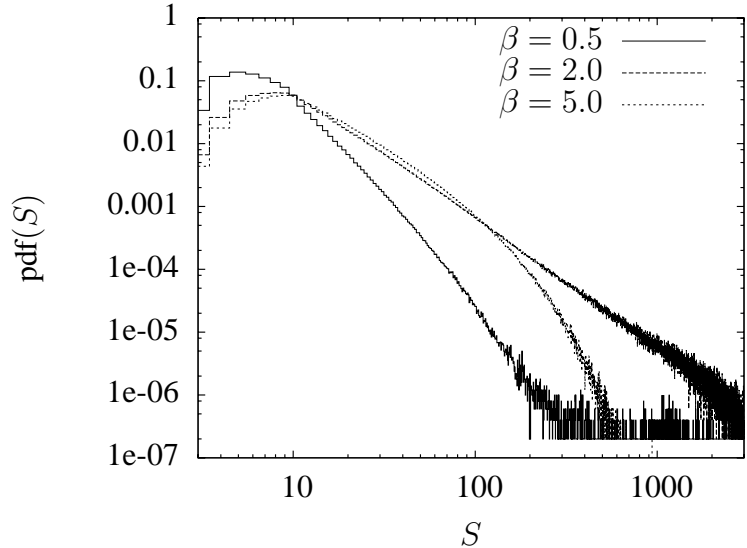}}
\caption{
Distribution of firm sizes plotted for different $\beta$. }
\label{fig3}
\end{figure}



\section{Conclusions}

By focusing on the geometric representation of firms
we proposed a stylized multi-agent model of firm growth. 
The competitive dynamics of firms under which the system 
reaches a steady state results a complex patterns of firm locations.
Despite of its formal simplicity, the model supplemented by the principle of minimum 
firm-firm correlation is able to explain the origin of the Pareto's law. 
Further validating of model is planned that takes into account 
real-world data. Hoverer, this will need to take into account the non-uniform distribution of customers.
The advanced model of this type is under development.

\paragraph{Acknowledgments}

The authors would like to express their 
thanks to Slovak Grant 
agency LPP APVV (grant no. 0098-06), VEGA (grant no. 1/4021/07, 1/2009/05) 
for financial support.  
\bibliographystyle{ICCS}
\bibliography{biblio}
\end{document}